\documentclass[fleqn,usenatbib]{mnras}

\usepackage{newtxtext,newtxmath}

\usepackage[T1]{fontenc}
\usepackage{ae,aecompl}

\usepackage{graphicx}	            
\usepackage{amsmath}	            
\usepackage{amssymb}	            
\usepackage{caption}                
\usepackage{pdfpages}               
\usepackage{float}                  
\usepackage{longtable}
\usepackage{lscape}

\title[Rotation measures of pulsars using CHIME/Pulsar]{Faraday rotation measures of northern-hemisphere pulsars using CHIME/Pulsar}

\author[Ng et al.]{
C. Ng$^{1}$,
A. Pandhi$^{2}$,
A. Naidu$^{3,4}$,
E. Fonseca$^{3,4}$,
V.~M.~Kaspi$^{3,4}$,
K.~W.~Masui$^{5,6}$,
\newauthor
R.~Mckinven$^{1,2}$,
A. Renard$^{1}$,
P.~Scholz$^{1,7}$,
I.~H.~Stairs$^{8}$,
S.~P.~Tendulkar$^{3,4}$,
K. Vanderlinde$^{1,2}$
\\
$^{1}$Dunlap Institute for Astronomy and Astrophysics, University of Toronto, 50 St. George Street, Toronto, ON M5S 3H4, Canada\\
$^{2}$David A. Dunlap Institute Department of Astronomy \& Astrophysics, University of Toronto, 50 St. George Street,
Toronto, ON M5S 3H4, Canada\\
$^{3}$Department of Physics, McGill University, 3600 rue University, Montr\'{e}al, QC H3A 2T8, Canada\\
$^{4}$McGill Space Institute, McGill University, 3550 rue University, Montr\'{e}al, QC H3A 2A7, Canada\\
$^{5}$MIT Kavli Institute for Astrophysics and Space Research, Massachusetts Institute of Technology, 77 Massachusetts Ave, Cambridge, MA 02139, USA\\
$^{6}$Department of Physics, Massachusetts Institute of Technology, Cambridge, 77 Massachusetts Ave, Massachusetts 02139, USA\\
$^{7}$Dominion Radio Astrophysical Observatory, Herzberg Astronomy \& Astrophysics Research Centre, National Research Council Canada, P.O. Box 248, Penticton, V2A 6J9, Canada\\
$^{8}$Dept. of Physics and Astronomy, University of British Columbia, 6224 Agricultural Road, Vancouver, BC V6T 1Z1, Canada
}

\date{Accepted XXX. Received YYY; in original form ZZZ}

\pubyear{2020}

\begin{document}
\label{firstpage}
\pagerange{\pageref{firstpage}--\pageref{lastpage}}
\maketitle

\begin{abstract}
Using commissioning data from the first year of operation of the Canadian 
Hydrogen Intensity Mapping Experiment's (CHIME) Pulsar backend system, 
we conduct a systematic analysis of the Faraday Rotation Measure (RM) of the northern hemisphere pulsars detected by CHIME. 
We present 55 new RMs as well as obtain improved RM uncertainties for 25 further pulsars. 
CHIME's low observing frequency and wide bandwidth between 400-800\,MHz contribute to the precision of our measurements, whereas the high cadence observation provide extremely high signal-to-noise co-added data. 
Our results represent a significant increase of the pulsar RM census, particularly regarding the northern hemisphere. 
These new RMs are for sources that are located in the Galactic plane out to 10\,kpc, as well as off the plane to a scale height of $\sim$16\,kpc.
This improved knowledge of the Faraday sky will contribute to future Galactic large-scale magnetic structure and ionosphere modelling. 
\end{abstract}


\section{Introduction}

Radio pulsars have emission that is among the most polarized of all astronomical sources.
As the linearly polarized emission travels through 
the interstellar medium (ISM) in the presence of magnetic fields, 
free electrons along the line of sight (LOS) lead to Faraday rotation of the observed pulsar radio emission. The differential of Faraday rotation (or the change in polarization angle, $\Delta$PA) between two observing frequencies is associated to the Rotation Measure (RM) and the observing wavelength ($\lambda$):
\begin{equation}
\Delta{\rm{PA}}={\rm{RM}}\lambda^{2}\,. \label{eq:pa}
\end{equation}
RM is in turn related to the number density of the free electrons along the LOS and the magnetic field of the LOS plasma ($B$) by:
\begin{equation}
{\rm{RM}}= 0.812\,{\rm rad}\,{\rm m}^{-2}\int_{0}^{d}\left[\frac{n_\text{e}(s)}{\text{cm}^{-3}}\right]\left[\frac{B(s)}{\mu \text{G}}\right] \cdot \left(\frac{ds}{\text{pc}}\right)\label{eq:RM} \,,
\end{equation}
where $d$ is the distance to the pulsar. At the same time, the pulsar emission is dispersed by the plasma electrons. The 
ratio of the RM to the dispersion measure (DM) 
provides an estimate of the integrated magnetic field strength along the parallel LOS:
\begin{equation}
\left<B_{\parallel}\right>=1.232\,\mu \text{G}\left(\frac{\text{RM}}{\text{rad}\, \text{m}^{-2}}\right)\left(\frac{\text{DM}}{\text{pc}\,\text{cm}^{-3}}\right)^{-1}\label{eq:B} \,.
\end{equation}

The Faraday rotation of pulsars is thus a powerful tool for studying the Galactic magnetic field.
The study of magnetic field structure is  important as 
it plays a critical role in numerous astrophysical processes, 
see for example, a review by \citet{Noutsos2012}.
Currently in the \textit{ATNF Pulsar Catalogue}\footnote{http://www.atnf.csiro.au/research/pulsar/psrcat} \citep[V1.61;][]{PSRCAT}, there are 1167 pulsars with a published RM, which is approximately 42\% of the full pulsar population. These pulsars are distributed throughout the Galactic disc, mapping the $\left<B_{\parallel}\right>$ of over a thousand LOS. Numerous previous studies have used pulsar RMs to model the large-scale component of the Galactic magnetic field  \citep[GMF;][]{Noutsos2015,Han2018,Sobey2019}. Pulsar RMs can also be combined with extragalactic RMs for a more comprehensive view of the Galactic disc and halo \citep[e.g.][]{vanEck2011}.

There is room for improvement in using pulsar RMs to model the GMF. 
Currently, a large fraction of the available pulsar RMs are located near the Galactic plane and are concentrated within a few kiloparsec from the Sun. Most of the previously measured  RMs were obtained with the Australian Parkes telescope in the southern hemisphere which operates at 1.4\,GHz. Since Faraday rotation depends on wavelength squared (Eq.~\ref{eq:pa}), lower frequency observations lead 
to more precise measurements of RMs.
Multiple aperture array telescopes at low frequency have come online in recent years. They will be providing an enhanced view of the polarized sky. These facilities include the Low-Frequency Array \citep[LOFAR;][]{vanHaarlem2013}, the Long-Wavelength Array \citep[LWA;][]{Taylor2012}, the Murchison Widefield Array \citep[MWA;][]{Tingay2013,Wayth2018}.
Most recently, the Canadian Hydrogen Intensity Mapping Experiment \citep[CHIME;][]{chimefrb}, which is a dense-packed interferometer operating at 400-800\,MHz, has also begun commissioning observations.

In this paper, we present a study of pulsar RMs using CHIME data. An overview of the telescope and observational data is provided in Section~\ref{sec:telescope} and the method of analysis is detailed in Section~\ref{sec:analysis}. We report on our results in Section~\ref{sec:results}, including 55 new RMs as well as improvement in the uncertainties of 25 catalogue RM values. In Section~\ref{sec:discuss}, we discuss the implications of our work and conclude in Section~\ref{sec:conclusion}.

\section{Telescope Overview and Observations}\label{sec:telescope}
\subsection{The CHIME/Pulsar backend}
CHIME is a radio telescope hosted by the Dominion Radio Astrophysical Observatory (DRAO) in British Columbia, Canada. 
CHIME operates across a wide bandwidth of 400$-$800\,MHz and has a collecting area ($\sim$80$\times$100\,m$^{2}$) and point-source sensitivity comparable to that of other 100-m class radio telescopes. 
The reflecting surface of CHIME consists of four parabolic cylinders. It is a transit telescope with no moving parts. For the CHIME/Pulsar project, we combine the signals from the 1024 dual polarization feeds and form 10 tied-array beams that are available as raw voltages \citep{Ng2018}. This means we can track 10 different pulsars at any given time as they transit through 
CHIME's field-of-view, along the meridian. 
This provides very high cadence scheduling: while many of the northern hemisphere pulsars are being monitored daily, the longest cadence to cycle through all sources in the northern sky is only $\sim$10 days. This is reflected in the long co-added integration length of our data (Total$_{\mathrm{fold}}$) and the high signal-to-noise (S/N) achieved as listed in Table~\ref{tab:RM}. 
The transit time of each source is a function of the declination; transit times can range from tens of minutes to hours for circumpolar sources. CHIME can in principle observe down to a declination of $-20^{\circ}$.

\subsection{Observations and Data Types}
The beamformed baseband data of the CHIME/Pulsar backend are dual-polarization, complex-sampled and split into 1024 frequency channels with 8 bits per complex sample.
For each observing scan, we generate `fold mode' archives using the `Digital Signal Processing for Pulsars' ({\tt dspsr}) suite, an open source GPU-based library developed by \citet{DSPSR}. These data are coherently dedispersed at the known DM (taken from the ATNF catalogue) and then folded at the catalogue spin period with nominally 256 phase bins at 10\,s sub-integration and the four Stokes ($I$, $Q$, $U$, and $V$) parameters are recorded. While 256 phase bins should be adequate for most of the pulsars studied in this work, we note that using too few phase bins could potentially lead to depolarization. A number of our pulsar observations were recorded with up to 2048 phase bins. For these sources, we step through the range of 32 to 2048 phase bins and see no obvious trend in the resulting RMs that is greater than their uncertainties.

The data have not yet been polarization calibrated. The beam shape of any aperture array is complicated, especially in the case of CHIME where over a thousand analog input components are involved. 
A proper calibration will be implemented in the near future.
Fortunately, Faraday rotation is generally unaffected because there are no wavelength-squared dependencies in the beam shape. We have verified our RM analysis pipeline by cross checking with 100 pulsars with catalogue RMs and obtained consistent results for most of them (see Section~\ref{sec:results-compare}). We note that uncalibrated data could potentially show instrumental polarization, where emission from Stokes $I$ contaminates the other Stokes parameters. This is manifested as a peak in the Faraday spectrum centred at zero; see for example, PSRs~J0026+6320, J0324+5239,  B0331+45, J0426+4933 in Fig~\ref{fig:plot1}. For a discrete sampled Faraday dispersion function, the FWHM of the theoretical RM spread function (RMSF) is given by $\delta\phi=3.8/\Delta(\lambda^2)$ \citep{Brentjens2005}. For CHIME, we have a $\delta\phi$ of about 9\,rad\,m$^{-2}$. The majority of our RM values have a higher magnitude than this so the presence of leakage does not cause any confusion. However, this could have an effect on small RMs and is further discussed in Section~\ref{sec:results-new}.

\section{Data reduction and analysis} \label{sec:analysis}
Our offline data analysis pipeline is heavily based on modules from the {\tt psrchive} pulsar data processing \citep{Hotan2004,vanStraten2012}. First, data affected by radio frequency interference (RFI) are excised by identifying outlier intensity values both in time and frequency. These sub-integrations and frequency channels are set to zero using {\tt paz} and {\tt pazi}. We create updated pulsar ephemerides using the {\tt tempo2} software package \citep{Hobbs2006} and we fit for DM using {\tt pdmp} in order to properly co-add all available fold mode data to obtain the highest possible S/N. The S/N-optimized DM for each pulsar is listed as DM$_\mathrm{obs}$ in Table~\ref{tab:RM}. We have not taken into account any temporal DM variations in this work. DM variations up to the order of 10$^{-2}$\,cm$^{-3}$pc/yr have been reported in the literature \citep[see, e.g.][]{Lam2015}. We have verified that offsetting the folding DM by this amount does not lead to difference in RM greater than the RM uncertainty quoted here.

We quantify the RM of these co-added data using the {\tt rmfit}\footnote{http://psrchive.sourceforge.net/manuals/rmfit/} tool. An initial guess of the RM is found by brute force searching the RM range of $\pm${1500}\,rad\,m$^{-2}$.
At each trial RM, {\tt rmfit} corrects for the associated Faraday rotation and computes the total linear polarization $L=(Q^{2}+U^{2})^{1/2}$ across the on-pulse region. {\tt rmfit} then fits a Gaussian to the resultant RM spectrum to identify a peak RM. 
This RM value is then iteratively refined as follows: the data are split in two equal frequency bands where each data segment is integrated and then compared to compute a weighted differential $\Delta$PA. The uncertainty of $\Delta$PA is minimized to obtain the best-fit RM value. Previous studies such as \citet{Morello2020} have shown that the iterative method of {\tt rmfit} can sometimes under-estimate the RM uncertainty. For pulsars that have high enough S/N for RM to be obtained on a per-session basis (i.e., without having to co-add across multiple days), we use the spread of one standard deviation of these daily RM values as our uncertainty. 
This applies to 39 out of the 80 updated RMs reported in this work. For the remaining pulsars which are too weak to obtain per-session RM, we adopt the RM uncertainty given by the brute-force method of {\tt rmfit} on the co-added data, which is conservatively calculated to be the range where the Gaussian fit for the RM spectrum drops by 2-$\sigma$.

Unmodelled profile evolution could lead to the addition of sine waves with amplitudes that have different frequency dependence and  potentially introducing bias in the RM measurement. Profile evolution is expected to be most rapid at lower observing frequencies below 200\,MHz \citep[see, e.g.][]{Phillips1992}. Upon a visual inspection, we do not see substantial variations among the pulsars in this sample and so have not attempted to fit for 2-D pulse profiles.
The effect of ionospheric Faraday rotation has not been corrected in this work. 
To do that properly requires a careful modelling of the ionosphere and is out of scope for this project. 
However, all the observations on which this paper is based were obtained during the current minimum of solar activity. We estimate that ionospheric RM was usually less than 1\,rad\,m$^{-2}$, and no more than $+$2\,rad\, m$^{-2}$ even during daytime \citep{Mevius2018a,Mevius2018b}.

\section{Results}\label{sec:results}

\subsection{Comparison to catalogue RMs} \label{sec:results-compare}
\begin{figure}
    \centering
    \includegraphics[scale=0.35]{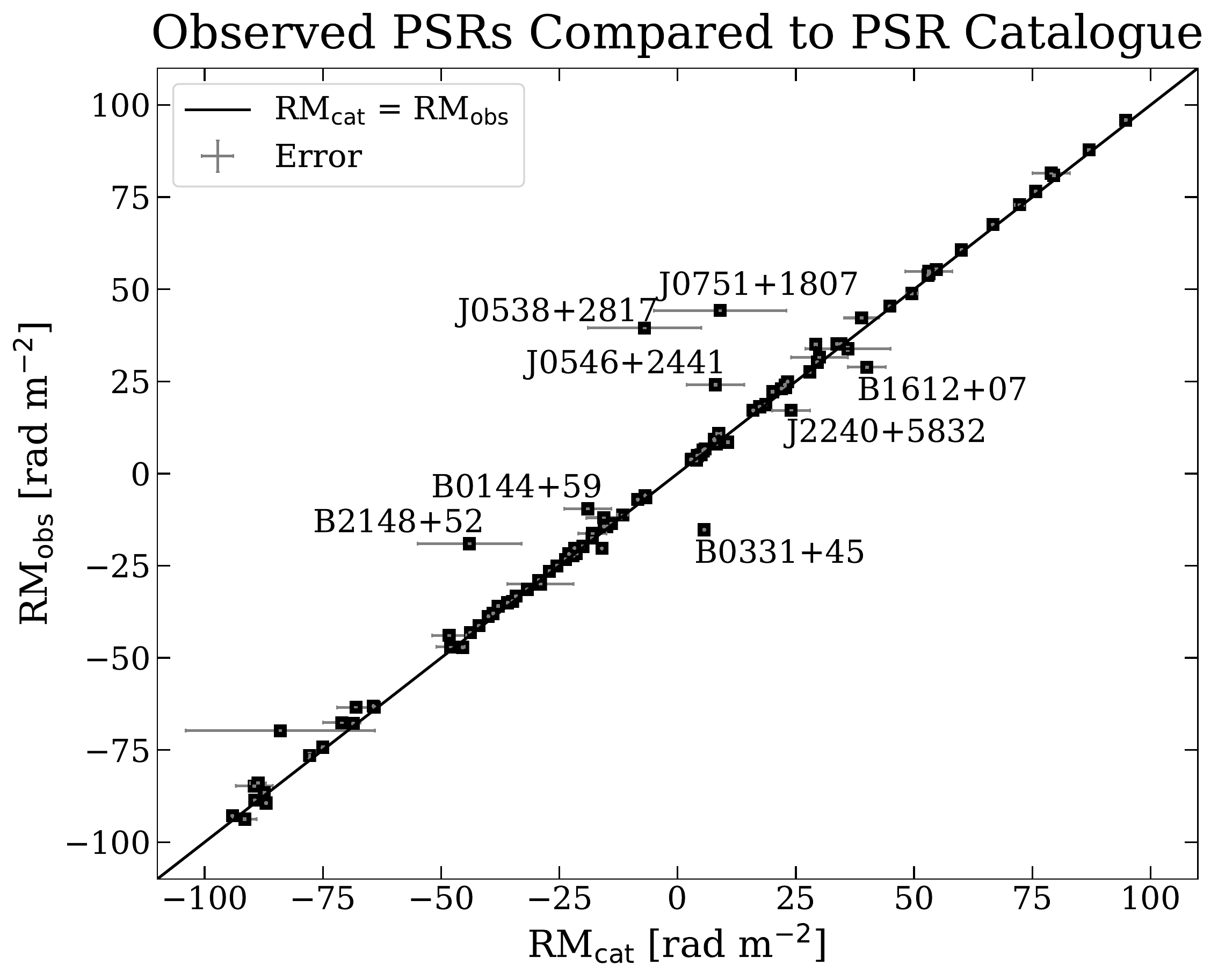}
    \caption{A comparison between RM$_{\mathrm{obs}}$ and RM$_{\mathrm{cat}}$ for 100 pulsars above $\delta=0^{\circ}$ detected by CHIME. The black line corresponds to the RM$_{\mathrm{obs}}$ = RM$_{\mathrm{cat}}$ trend which we expect to see. PSRs~B0144+59, B0331+45, J0538+2817, J0546+2441, J0751+1807, B1612+07, B2148+52 and J2240+5832 are highlighted as they significantly deviate from the RM$_{\mathrm{obs}}$ = RM$_{\mathrm{cat}}$ trend; these eight pulsars are discussed further in the main text.}
    \label{fig:compare}
\end{figure}

The data set used in this work is from the first year of CHIME/Pulsar commissioning observations, which spans September 2018 to 2019.
In order to verify our RM pipeline and the CHIME data, we first attempt to cross check RMs for pulsars that have a catalogue RM value. 
We analyse all pulsars above declination $\delta=0^{\circ}$ detected by CHIME that have a published RM between $\pm$100\,rad\,m$^{-2}$. We detect unambiguous RMs in 100 pulsars and use them as our verification sample.
As can be seen from Fig.~\ref{fig:compare}, most of our observed RM (RM$_{\mathrm{obs}}$) agree with the catalogue value (RM$_{\mathrm{cat}}$), which is reassuring. 
In addition, we measure RMs with substantially lower uncertainties than those in the ATNF pulsar catalogue (V1.61) for 25 pulsars.
 This is likely due to the higher precision RM made possible from our lower frequency and large bandwidth observations, as well as the high S/N profiles, which result from long-duration co-added data. 

Eight of the RMs we measure differ significantly from their  catalogue values. These sources are annotated in Fig.~\ref{fig:compare}.
The  catalogue  RM  of  three  of  those,  namely PSRs~J0538+2817,
J0546+2441, and J0751+1807, are close to zero, which suggests that previous studies might have been contaminated by spectral
leakage.
We find that the RMs of four pulsars have changed by a few to a few tens of RM units in the time between their catalogue measurement epochs and our more recent CHIME detections. 
These include  
PSR~B0144+59 with 
RM$_{\mathrm{cat}}$=$-$19(4)\,rad\,m$^{-2}$ \citep{Rand1994} compared to  RM$_{\mathrm{obs}}$=$-$9.5(5)\,rad\,m$^{-2}$, 
PSR~B1612+07 with 
RM$_{\mathrm{cat}}$=40(4)\,rad\,m$^{-2}$ \citep{Hamilton1987} compared to  RM$_{\mathrm{obs}}$=28(3)\,rad\,m$^{-2}$, 
PSR~B2148+52 with 
RM$_{\mathrm{cat}}$=$-$44(11)\,rad\,m$^{-2}$ \citep{Mitra2003} compared to  RM$_{\mathrm{obs}}$=$-$19.0(8)\,rad\,m$^{-2}$, and 
PSR~J2240+5832 with 
RM$_{\mathrm{cat}}$=24(4)\,rad\,m$^{-2}$ \citep{Theureau2011} compared to 
RM$_{\mathrm{obs}}$=17.1(11)\,rad\,m$^{-2}$. 
These differences are too big to be explained by any ionospheric corrections. 
It is possible that we are seeing a genuine temporal evolution of these RMs over the years, although there are no other data points in the literature to help verify this. The presence of an unusual local environment (e.g. the Supernova Remnant of Vela \citep{Johnston2005}, an eclipsing black widow binary \citep{Breton2013}) has been suggested to cause
a change in temporal RM, although these scenarios do not seem to apply to the four pulsars listed above. 
Finally, for PSR~B0331+45 we obtain an RM$_{\mathrm{obs}}$ of $-$15.2(11)\,rad\,m$^{-2}$ which is discrepant with the latest result from \citet{Sobey2019} at 5.60(9)\,rad\,m$^{-2}$. Even though our data of PSR~B0331+45 is affected by instrumental leakage which shows up as a peak near zero in the RM spectrum (see Fig.~\ref{fig:plot1}), we detect a clear non-zero RM and the signal is confirmed upon inspecting the expected oscillation in Stokes~$Q$. Given that our work and that of \citet{Sobey2019} are taken relatively close in time, it seems unlikely that the RM could have changed so much within a year or so. We note that there is an older published RM of this pulsar at $-$41(20)\,rad\,m$^{-2}$ by \citet{Rand1994}, which is consistent with our observed value. 

\subsection{New RMs for 55 pulsars} \label{sec:results-new}
Over 500 known pulsars were detected by CHIME/Pulsar above $\delta=-20^{\circ}$ in the first year of commissioning. 
We have sufficiently high S/N data for RM measurement for 109 of these pulsars that do not already have published RM values.
We obtain new RMs for 55 of these pulsars; 
see Table~\ref{tab:RM} for details, and 
Figs.~\ref{fig:plot1}-\ref{fig:plot3}
for the individual RM spectra.
We are not able to quantify RMs for the remaining pulsars because of two reasons: (1) instrumental leakage dominates 30 of those, which could potentially be improved when a  polarization calibration scheme is carried out on CHIME data in the future; (2) no RM peak is detected in 24 pulsars, which could be because these pulsars are intrinsically weakly polarized. Refer to 
Appendix~\ref{sec:noRM} for a list of these pulsars.

Thirteen of these 55 pulsars were already studied using LOFAR in \citet{Sobey2019}, although they did not detect any significant RM at the time (refer to Table~2 in that publication). 
This may be because LOFAR's frequency channel widths are sufficiently broad that they lose >50\% sensitivity for
|RM|>163\,rad\,m$^{-2}$, whereas for CHIME/Pulsar, with our frequency resolution of 390\,kHz, we maintain full sensitivity out to a RM of roughly 1000\,rad\,m$^{-2}$. Some of these pulsars lie along the Galactic plane, and LOFAR could be suffering from more scattering and depolarization at the high DM and RM due to its low observing frequency. We note that there is likely some degree of profile scattering for some of the pulsars in the CHIME band as well. We have not attempted to model the effect of scattering on RM here; the difference is expected to be small and within the uncertainty of our RM values. Finally, 
the remarkably long duration from the co-addition of data significantly increases our ability to conduct this kind of analysis on a large number of pulsars that would otherwise not have the required S/N to robustly measure the RM.

As mentioned before, the data configuration of 
CHIME implies that we have a theoretical RMSF of $\sim$9\,rad\,m$^{-2}$. 
Three of our new RMs fall within this range, notably PSR~ J1647+6608 with RM$_{\mathrm{obs}}$=7(3)\,rad\,m$^{-2}$, 
PSR~J1911+1347 with RM$_{\mathrm{obs}}$=$-$7(3)\,rad\,m$^{-2}$, and PSR~J2340+08 with RM$_{\mathrm{obs}}$=$-$7(2)\,rad\,m$^{-2}$. Judging from their respective RM spectrum (see Fig.~\ref{fig:plot1}-\ref{fig:plot3}), the RM peak is clearly distinct from zero. This made us believe that these new RMs are reliable and not due to confusion with instrumental leakage. 

\section{Discussion} \label{sec:discuss}

We have updated 25 existing RM values of northern hemisphere pulsars and measured 55 new RMs. 
These 80 pulsars are located in our full range of Right Ascension (RA) and $\delta$, spanning  1.5$^{\circ}\leq$RA$\leq355.8^{\circ}$ and $-6.7^{\circ}\leq\delta\leq83.2^{\circ}$.
 In terms of Galactic coordinates, which provide useful insight for studies looking at the Milky Way's magnetic field along longitudinal LOS, the pulsars lie within Galactic longitude of $20.6^{\circ}\leq l \leq 219.4^{\circ}$ and Galactic latitude of $-50.4^{\circ}\leq b \leq54.2^{\circ}$. 
The range of updated RMs reported in this work lies between $-$295\,rad\,m$^{-2}\leq$RM$_{\mathrm{obs}}\leq$338\,rad\,m$^{-2}$. The average $|$RM$_{\mathrm{obs}}|$ in this study is roughly 70\,rad\,m$^{-2}$. 
Combining the DM$_{\mathrm{obs}}$ listed in Table~\ref{tab:RM}, we can apply Equation~\ref{eq:B} to calculate the LOS parallel magnetic field strength, $\left<B_{\parallel}\right>$ (last column of Table~\ref{tab:RM}). According to \citet{Sun2010}, the strength of the regular halo magnetic field is about 2\,$\upmu$G, with which our results are largely consistent. 
 
 If we take into account the DM-derived distance of these pulsars as listed in the ATNF Pulsar Catalogue \citep[based on the electron density model of][]{YMW16}, we can locate the derived $\left<B_{\parallel}\right>$ on a 3-D plot in Cartesian coordinates, as shown in Figs.~\ref{fig:xyproj}-\ref{fig:yzproj}. Note that most of the known pulsars do not have independent distance measurements (e.g. from parallax), and a majority of the pulsar distances are obtained from their DM, combined with a model of the free electron distribution, which is said to have uncertainty up to some 20\%. Keeping the caveat of the distance uncertainty in mind, we can still form an overview picture of the updated Faraday sky. 
  In Fig.~\ref{fig:xyproj}, we overlay the location of RM values from this work as well as all existing catalogue pulsar RMs with a simulation of the spiral arm structure of our Milky Way. It can be seen that most of our RMs fall within quadrants III and IV, and they also somewhat follow the arm structure.
 Compared to the recent LOFAR study by \citet{Sobey2019}, we typically measure RMs farther out along the Galactic plane (up to $\sim$10\,kpc) whereas LOFAR's sample is mostly concentrated near the Sun. This is likely due to less severe scattering and depolarization at the higher observing frequency of CHIME. 
 In terms of scale height ($z$), we cover a large range, detecting RMs up to $z\sim$16\,kpc.
 From our 80 updated RMs, we see a similar dichotomy already mentioned by \citet{Sobey2019}, that $\left<B_{\parallel}\right>$ values above the Galactic plane tend to be positive while those below the plane tend to be negative. 

\section{Conclusions} \label{sec:conclusion}
We present new measurements of RMs for northern radio pulsars, using commissioning data from CHIME/Pulsar. We report 55 new RMs
, and provide improved values and uncertainties for 25 pulsars with previously catalogued RMs. 
The wide bandwidth of CHIME 
has enabled our high RM precision.  The observing frequency of 400-800\,MHz is at a sweet spot for RM studies, low enough to take advantage of the wavelength squared dependency of RM and the steep spectrum of most pulsars, but not so low as to be hindered by scattering and depolarization at high DM and RM. The high cadence observations of CHIME also provide excellent S/N in our co-added data. We note that ionospheric RM corrections have not been applied in this work and it remains the biggest source of systematic uncertainty in our RMs. 

The 80 updated RMs reported here cover a large region of the Milky Way, both deep in the Galactic plane and far out in the halo. The derived $\left<B_{\parallel}\right>$ values are comparable to the average magnetic field strength of our Galaxy. 
Overall, this work improves our knowledge of the Faraday sky, with potential implications on future Galactic large-scale structure and ionospheric modelling studies.

\section*{Acknowledgements}
We are grateful to the staff of the Dominion Radio Astrophysical Observatory, which is operated by the National Research Council Canada.
Pulsar research at UBC is supported by an NSERC Discovery Grant and by the Canadian Institute for Advanced Research.
CN is a SOSCIP TalentEdge fellow.
AP was supported by the Summer Undergraduate Research Program (SURP) in astronomy \& astrophysics at the University of Toronto.
V.M.K. holds the Lorne Trottier Chair in Astrophysics \& Cosmology and a Canada Research Chair and receives support from an NSERC Discovery Grant and Herzberg Award, from an R. Howard Webster Foundation Fellowship from the Canadian Institute for Advanced Research (CIFAR), and from the FRQNT Centre de Recherche en Astrophysique du Quebec.
P.S. is a Dunlap Fellow and an NSERC Postdoctoral Fellow. The Dunlap Institute is funded through an endowment established by the David Dunlap family and the University of Toronto. 
We thank Tom Landecker for useful discussion and Bradley Meyers for carefully reading the manuscript. We also thank Jumei Yao for providing the electron density data for the YMW2016 model. 

\bibliographystyle{mnras}
\bibliography{ref}


\appendix

\section{Summary of Results} \label{app:table}
Summarized in Table~\ref{tab:RM} are the principal results of this study using the {\tt rmfit} tool on co-added pulsar observations from CHIME/Pulsar. This Table comprises 80 pulsars: 55 new RM results that have not been previously published and 25 updated RMs to the pulsar catalogue (version 1.61) with tighter bounds on the uncertainty.

Column~1 shows the pulsar name in B1950 or J2000 coordinate systems; the pulsars are listed in order of ascending RA. Columns~2$-$3 show the published RM results and uncertainties from the latest version of the ATNF pulsar catalogue (V~1.61) and column~4 includes a short-hand of the corresponding literature reference in which the result was published. 
The full citation of these entries can be found in Table~\ref{tab:ref}.
Asterisks in columns~2$-$4 denote results that do not have a previously published RM result in the pulsar catalogue. Column~5 lists the MJD range of our CHIME/Pulsar data set and column~6 indicates the total length of fold mode observation in the co-added data. The S/N of the co-added profile can be found in column~7. The DM and the RM obtained from the co-added data are tabulated in columns 8$-$11. 
The uncertainty of our observed RM comes from the spread of the per-session RMs. When that is not available for the low S/N pulsars, uncertainty is taken from the RM spectrum Gaussian fit width calculated by {\tt rmfit}. Our RMs do not include any ionospheric corrections. 
Columns~12$-$13 list the $\left<B_{\parallel}\right>$ and associated error derived from applying Equation~\ref{eq:B} to our DM and RM results. 

\onecolumn
\begin{landscape}
\begin{center}
\begin{longtable}{ccccccccccccc}
\caption{Summary of 80 co-added RMs obtained in this work. See Appendix~\ref{app:table} for a detailed description of each column.} \label{tab:RM} \\
\hline 
PSR & RM$_{\mathrm{cat}}$ & $\pm$ & Ref & MJD & Total$_{\mathrm{fold}}$ & S/N & DM$_{\mathrm{obs}}$  & $\pm$ & RM$_{\mathrm{obs}}$ & $\pm$ & $\left<B_{\parallel}\right>$ & $\pm$\\
    & (rad\,m$^{-2}$) & (rad\,m$^{-2}$) &    &  & (hrs) & & (pc\,cm$^{-3}$) & (pc\,cm$^{-3}$) & (rad\,m$^{-2}$) & (rad\,m$^{-2}$) & ($\upmu$G) & ($\upmu$G)\\
\hline
\endfirsthead

\multicolumn{6}{c}%
{{\bfseries \tablename\ \thetable{} -- continued from previous page}} \\
\hline
  PSR & RM$_{\mathrm{cat}}$ & $\pm$ & Ref & MJD & Total$_{\mathrm{fold}}$ & S/N & DM$_{\mathrm{obs}}$  & $\pm$ & RM$_{\mathrm{obs}}$ & $\pm$ & $\left<B_{\parallel}\right>$ & $\pm$ \\
  & (rad\,m$^{-2}$) & (rad\,m$^{-2}$) &    &  & (hrs) & & (pc\,cm$^{-3}$) & (pc\,cm$^{-3}$) & (rad\,m$^{-2}$) & (rad\,m$^{-2}$) & ($\upmu$G) & ($\upmu$G)\\
\hline
\endhead

\hline \multicolumn{2}{c}{{Continued on next page}} \\ \hline
\endfoot
\hline \hline
\endlastfoot
  J0006+1834& *       & *     & *      & 58441-58740 & 10.7 & 76   & 12.4606 & 0.2000 & $-$20    & 3   & $-$2.008& 0.041\\
  J0026+6320& *       & *     & *      & 58441-58467 & 1.2  & 82   & 245.1196& 0.1000 & $-$294   & 2   & $-$1.483& 0.002\\
  J0058+6125& *       & *     & *      & 58441-58742 & 38.0 & 94   & 128.6608& 0.2000 & $-$212   & 2   & $-$2.032& 0.006\\
  B0138+59  & $-$48.00& 3.00  & hl87   & 58520-58711 & 38.4 & 1752 & 34.4630 & 0.4000 & $-$47.0  & 1.1 & $-$1.684& 0.024\\
  B0144+59  & $-$19.00& 5.00  & rl94   & 58321-58720 & 34.2 & 811  & 40.0862 & 0.0700 & $-$9.5   & 0.5 & $-$0.294& 0.004\\
  B0154+61  & $-$29.00& 7.00  & hl87   & 58520-58701 & 19.4 & 735  & 29.9132 & 9.0000 & $-$29    & 3   & $-$1.272& 0.457\\
  J0201+7005& *       & *     & *      & 58634-58732 & 42.0 & 144  & 20.8587 & 0.5000 & $-$39    & 3   & $-$2.320& 0.071\\
  J0324+5239& *       & *     & *      & 58441-58692 & 15.9 & 188  & 115.4211& 0.1000 & $-$126   & 3   & $-$1.348& 0.003\\
  B0331+45  & 5.60    & 0.09  & sbg+19 & 58398-58720 & 14.6 & 845  & 47.1117 & 0.1000 & $-$15.2  & 1.1 & $-$0.398& 0.003\\
  B0339+53  & $-$84.00& 20.00 & mwkj03 & 58351-58715 & 17.6 & 606  & 68.0324 & 0.7000 & $-$69.7  & 1.5 & $-$1.264& 0.016\\
  J0340+4130& *       & *     & *      & 58520-58744 & 81.9 & 361  & 49.5823 & 0.0003 &   56.1   & 0.7 &    1.395& 0.002\\
  B0355+54  & 79.00   & 4.00  & hl87   & 58318-58711 & 41.0 & 6561 & 57.1223 & 0.0600 &    81.5  & 0.3 &    1.758& 0.003\\
  J0426+4933& *       & *     & *      & 58500-58691 & 6.5  & 346  & 84.1850 & 0.3000 & $-$169   & 7   & $-$2.485& 0.012\\
  J0453+1559& *       & *     & *      & 58439-58688 & 7.6  & 103  & 30.2880 & 0.0100 & $-$35    & 2   & $-$1.435& 0.018\\
  J0517+2212& $-$16.00& 0.00  & bfrs18 & 58439-58714 & 10.9 & 1030 & 18.6769 & 0.0800 & $-$20.2  & 1.6 & $-$1.338& 0.009\\
  J0538+2817& $-$7.00 & 12.00 & mwkj03 & 58321-58709 & 8.9  & 886  & 39.8826 & 0.0500 &    39    & 3   &    1.220& 0.002\\
  J0540+3207& *       & *     & *      & 58520-58701 & 19.2 & 370  & 61.9038 & 0.2000 &    13.7  & 1.8 &    0.273& 0.004\\
  J0546+2441& 8.00    & 6.10  & hmvd18 & 58320-58735 & 12.6 & 294  & 73.0922 & 1.0000 &    24    & 3   &    0.407& 0.008\\
  J0555+3948& *       & *     & *      & 58520-58744 & 19.6 & 183  & 36.7105 & 0.4000 &     9    & 3   &    0.325& 0.015\\
  J0613+3731& *       & *     & *      & 58518-58720 & 5.5  & 178  & 18.7556 & 0.2000 &    16    & 2   &    1.082& 0.017\\
  J0614+83  & *       & *     & *      & 58702-58737 & 19.2 & 45   & 43.8689 & 0.4000 & $-$13    & 3   & $-$0.385& 0.090\\
  J0627+0649& *       & *     & *      & 58456-58692 & 2.9  & 113  & 86.5563 & 0.1000 &   179    & 3   &    2.555& 0.004\\
  J0646+0905& *       & *     & *      & 58519-58640 & 8.7  & 190  & 148.8859& 0.3000 & $-$149   & 2   & $-$1.239& 0.003\\
  J0711+0931& *       & *     & *      & 58520-58744 & 8.5  & 127  & 46.0848 & 0.4000 &    62.8  & 1.1 &    1.684& 0.018\\
  J0740+6620& *       & *     & *      & 58517-58731 & 78.5 & 385  &14.9626  & 0.0002 & $-$36    & 2   & $-$3.040& 0.010\\
  J0751+1807& 9.00    & 14.00 & hmvd18 & 58321-58701 & 11.2 & 149  &30.2438  & 0.0010 &    44    & 3   &    1.803& 0.006\\
  J0815+0939& 53.10   & 5.00  & hmvd18 & 58321-58715 & 10.0 & 127  & 53.0660 & 0.2000 &    54    & 2   &    1.275& 0.009\\
  J0843+0719& *       & *     & *      & 58321-58715 & 9.4  & 48   & 33.5864 & 0.5000 &    40    & 4   &    1.506& 0.037\\
  J1434+7257& *       & *     & *      & 58516-58700 & 34.5 & 267  & 12.5997 & 0.0100 &  $-$9.7  & 1.2 & $-$0.951& 0.017\\
  J1518+4904& $-$15.60& 3.70  & hmvd18 & 58441-58733 & 17.6 & 2714 & 11.6062 & 0.0100 & $-$11.9  & 1.3 & $-$1.271& 0.013\\
  J1538+2345& *       & *     & *      & 58520-58741 & 9.7  & 141  & 14.4737 & 1.0000 &    11.5  & 1.1 &    1.020& 0.093\\
  J1544+4937& *       & *     & *      & 58520-58742 & 54.5 & 248  &23.2280  & 0.0008 &     9.8  & 1.9 &    0.523& 0.010\\
  B1612+07  & 40.00   & 4.00  & hl87   & 58329-58715 & 11.9 & 440  & 21.2426 & 0.4000 &    28    & 3   &    1.677& 0.042\\
  B1639+36A & *       & *     & *      & 58359-58693 & 25.3 & 112  & 30.4310 & 0.0040 &    13    & 3   &    0.532& 0.012\\
  J1647+6608& *       & *     & *      & 58501-58742 & 22.5 & 107  & 22.7981 & 0.6000 &     7    & 3   &    0.411& 0.017\\
  J1710+4923& *       & *     & *      & 58542-58721 & 13.2 & 193  & 7.0826  & 0.0010 &    12    & 2   &    2.117& 0.165\\
  J1736+05  & *       & *     & *      & 58450-58718 & 10.0 & 54   & 38.6724 & 0.3000 &    44    & 3   &    1.417& 0.027\\
  J1738+0333& 36      & 9     & gmd+18 & 58515-58744 & 35.3 & 70   &33.7652  & 0.0040 &    33    & 3   &    1.235& 0.027\\
  B1802+03  & 38.90   & 3.70  & hmvd18 & 58328-58711 & 39.2 & 285  & 80.8294 & 0.0800 &    42    & 2   &    0.644& 0.005\\
  J1821+0155& *       & *     & *      & 58683-58719 & 4.7  & 56   & 51.7499 & 0.0100 &   110    & 3   &    2.638& 0.014\\
  J1834+10  & *       & *     & *      & 58721-58739 & 2.0  & 23   & 78.0350 & 0.7000 &    97.0  & 1.7 &    1.533& 0.041\\
  J1900+30  & *       & *     & *      & 58715-58739 & 3.9  & 128  & 71.7592 & 0.2000 &   121    & 2   &    2.094& 0.011\\
  J1911+1347& *       & *     & *      & 58472-58742 & 24.6 & 87   &30.9806  & 0.0004 &   $-$7   & 3   & $-$0.282& 0.011\\
  J1918-0642& *       & *     & *      & 58500-58742 & 31.7 & 102  &26.5890  & 0.0030 & $-$57    & 3   & $-$2.648& 0.207\\
  B1926+18  & *       & *     & *      & 58365-58660 & 3.0  & 49   & 111.5379& 0.4000 &   174    & 2   &    1.931& 0.012\\
  J1946+2535& *       & *     & *      & 58443-58708 & 2.0  & 39   &248.4199 & 0.1000 &    57    & 3   &    0.293& 0.015\\
  J1946+2611& $-$88.70& 1.70  & hmvd18 & 58461-58662 & 3.3  & 93   & 165.3295& 0.1000 & $-$83.9  & 1.3 & $-$0.626& 0.002\\
  B1952+29  & $-$18.00& 3.00  & hl87   & 58501-58662 & 6.3  & 530  & 7.8782  & 0.1000 & $-$16    & 2   & $-$2.567& 0.041\\
  B2002+31  & 30.00   & 6.00  & hl87   & 58520-58701 & 32.4 & 1451 & 234.5535& 0.8000 &    31.5  & 0.8 &    0.166& 0.001\\
  J2010+2845& *       & *     & *      & 58450-58705 & 5.2  & 146  & 112.3273& 0.2000 & $-$233   & 3   & $-$2.560& 0.007\\
  J2011+3331& *       & *     & *      & 58521-58664 & 13.4 & 172  & 298.3448& 0.3000 &   235.6  & 0.5 &    0.973& 0.002\\
  J2013+3058& *       & *     & *      & 58441-58650 & 6.6  & 174  & 148.6652& 0.1000 & $-$145   & 3   & $-$1.204& 0.007\\
  J2016+1948& *       & *     & *      & 58451-58661 & 7.2  & 131  & 33.7492 & 0.2000 & $-$121   & 4   & $-$4.419& 0.033\\
  J2017+5906& *       & *     & *      & 58509-58636 & 10.6 & 174  & 59.9378 & 0.1000 &    33    & 3   &    0.696& 0.005\\
  J2019+2425& $-$71   & 4     & gmd+18 & 58521-58677 & 7.8  & 189  &17.1985  & 0.0010 & $-$67.6  & 0.6 & $-$4.843& 0.006\\
  B2025+21  & *       & *     & *      & 58448-58664 & 8.1  & 64   & 96.9408 & 0.1000 & $-$210   & 2   & $-$2.680& 0.007\\
  J2027+4557& *       & *     & *      & 58447-58717 & 8.5  & 368  & 229.4552& 0.4000 &   337    & 3   &    1.815& 0.004\\
  J2040+1657& *       & *     & *      & 58445-58737 & 5.1  & 72   & 50.2550 & 0.3000 & $-$98    & 2   & $-$2.413& 0.021\\
  J2045+0912& $-$89.50& 3.90  & hmvd18 & 58445-58737 & 21.7 & 338  & 31.3267 & 0.1000 & $-$84.7  & 1.9 & $-$3.337& 0.015\\
  J2047+5029& *       & *     & *      & 58487-58701 & 6.7  & 118  & 107.8448& 0.1000 & $-$121   & 4   & $-$1.394& 0.004\\
  B2053+36  & $-$68.00& 4.00  & hl87   & 58440-58700 & 43.2 & 3082 & 97.3875 & 0.0800 & $-$63.4  & 0.3 & $-$0.802& 0.001\\
  J2102+38  & *       & *     &  *     & 58440-58701 & 28.0 & 255  & 86.0513 & 0.4000 & $-$74    & 2   & $-$1.062& 0.007\\
  B2122+13  & $-$48.30& 3.60  & hr10   & 58440-58711 & 9.9  & 242  & 30.1597 & 0.2000 & $-$43    & 3   & $-$1.797& 0.016\\
  J2138+4911& *       & *     & *      & 58440-58693 & 7.3  & 441  & 168.2636& 0.2000 & $-$143   & 3   & $-$1.052& 0.002\\
  J2139+00  & *       & *     & *      & 58445-58693 & 27.6 & 353  & 31.2913 & 0.1000 &    11    & 2   &    0.439& 0.007\\
  B2148+52  & $-$44.00& 11.00 & mwkj03 & 58443-58715 & 57.5 & 1564 & 149.0138& 0.1000 & $-$19.0  & 0.8 & $-$0.157& 0.001\\
  J2208+4056& *       & *     & *      & 58452-58722 & 13.2 & 76   & 11.7588 & 0.2000 & $-$40    & 3   & $-$4.316& 0.106\\
  J2208+5500& *       & *     & *      & 58439-58693 & 5.1  & 234  & 104.5698& 0.3000 & $-$87    & 3   & $-$1.027& 0.007\\
  J2216+5759& *       & *     & *      & 58521-58722 & 12.7 & 184  & 167.3322& 0.1000 & $-$10.9  & 1.7 & $-$0.081& 0.001\\
  J2217+5733& *       & *     & *      & 58456-58718 & 16.0 & 491  &130.6711 & 0.4000 & $-$90    & 3   & $-$0.085& 0.004\\
  J2227+30  & *       & *     & *      & 58456-58605 & 5.1  & 589  & 19.9559 & 0.3000 & $-$58.3  & 1.9 & $-$3.612& 0.068\\
  J2229+6114& *       & *     & *      & 58451-58664 & 16.1 & 484  &204.9635 & 0.0200 & $-$187   & 3   & $-$1.129& 0.001\\
  J2234+2114& $-$91.50& 2.50  & hmvd18 & 58652-58722 & 3.9  & 191  & 34.7371 & 0.5000 & $-$93.7  & 1.8 & $-$3.329& 0.060\\
  J2240+5832& 24.00   & 4.00  & tpc+11 & 58455-58701 & 17.8 & 357  &263.4647 & 0.0500 &    17.1  & 1.1 &    0.080& 0.001\\
  J2243+1518& *       & *     & *      & 58440-58722 & 10.0 & 66   & 39.6448 & 0.2000 & $-$35.5  & 0.5 & $-$1.120& 0.020\\
  J2302+4442& *       & *     & *      & 58450-58709 & 45.0 & 269  &13.7196  & 0.0005 &    19.1  & 1.6 &    1.715& 0.004\\
  J2310+6706& *       & *     & *      & 58440-58651 & 3.7  & 29   & 98.5726 & 0.7000 & $-$16    & 2   & $-$0.204& 0.008\\
  J2319+6411& *       & *     & *      & 58519-58731 & 9.8  & 1393 &245.9455 & 0.0800 & $-$48.8  & 1.2 & $-$0.244& 0.001\\
  J2340+08  & *       & *     & *      & 58519-58701 & 8.8  & 118  & 23.9234 & 0.1000 &  $-$7    & 2   & $-$0.388& 0.022\\
  J2343+6221& *       & *     & *      & 58652-58701 & 65.7 & 303  & 155.8648& 0.6000 & $-$44.9  & 0.9 & $-$0.356& 0.002\\
  \hline
 \end{longtable}
 \end{center}
 \end{landscape}
\twocolumn

\begin{table}
\begin{center}
\caption{Summary of the references for the catalogue RMs listed in Table~\ref{tab:RM}.}
\label{tab:ref}
\begin{tabular}{cccc} 
\hline
\hline
Shorthand & Reference & (lowest) Centre frequency (MHz) & Ionospheric RM corrected \\
\hline
 bfrs18 & \cite{Brinkman2018} &  327 & N\\
 gmd+18 & \cite{Gentile2018} & 1400 & N \\
 hl87   & \cite{Hamilton1987} & 408  & Y\\ 
 hmvd18 & \cite{Han2018} & 774  & Y\\ 
 hr10   & \cite{Hankins2010} & 50  & N \\
 lbr+13 & \cite{Lynch2013} & 350/820 & N\\
 mwkj03 & \cite{Mitra2003} & 1400  & N\\
 rl94   & \cite{Rand1994}   & 1400 & N \\
 sbg+19 & \cite{Sobey2019}  & 110 & Y\\
 tpc+11 & \cite{Theureau2011} & 1100 & N\\
\hline
\end{tabular}
\end{center}
\end{table}

\section{Summary of Faraday Spectra Observations} \label{app:figure}
Figures~\ref{fig:plot1} through \ref{fig:plot3} show the individual Faraday spectra of the 80 RM results presented in Table~\ref{tab:RM}. The x-axes show the appropriate RM range that incorporates the peak RM values. The y-axes show normalized polarized flux in arbitrary units as our data are uncalibrated in polarization. Peaks centred at 0\,rad\,m$^{-2}$ and symmetric peaks about 0\,rad\,m$^{-2}$ are due to instrumental leakage discussed in Section~\ref{sec:telescope}. The RM$_{\mathrm{obs}}$ is shown as a vertical dashed line and its corresponding uncertainty is represented by vertical solid lines, although in almost all cases the uncertainty range is too small to visually separate the three lines. When available, the catalogue RM (RM$_{\mathrm{cat}}$) is represented by a vertical red dash-dot line and its corresponding uncertainty region is marked by the shaded hatch. 

\section{Pulsars with no detected RM} \label{sec:noRM}
In addition to the 80 RMs reported in this work, a further 54 pulsars were studied although no RM was detected in them. Instrumental leakage dominated 30 pulsars, namely PSRs~
J0023+0923, J0051+0423, J0243+6027, J0329+1654, J0458$-$0505, J0609+2130, J0611+1436, J0621+0336, J0630$-$0046, J0647+0913,
J1125+7819, J1327$-$0755, J1501$-$0046, J1628+4406, B1740$-$13,
J1744$-$1610, J1745$-$0129, J1802+0128, J1807+0756, B1810+02, 
J1820$-$0509, B1831$-$00, J2018+3431, J2048+2255, J2123+5434, 
J2206+6151, J2228+6447, J2325$-$0530, J2333+6145, and J2352+65. 
A further 24 pulsars likely have intrinsically weak polarization, namely PSRs~
J0103+54, J0139+5621, J0332+79, J0337+1715, J0645+80, J0652$-$0142,
B1740$-$03, J1819$-$1318, J1842+0638, J1843+2024, J1846$-$0749, J1846$-$07492, J1848+0826, B1904+12, B1906+09, J1925+19, J1931+30,
B1933+17, J1954+4357, B1957+20, J2002+30, J2013$-$0649, J2015+2524, 
and J2030+55.

\begin{figure*}
\centering
\setlength\fboxsep{0pt}
\setlength\fboxrule{0pt}
\fbox{\includegraphics[width=17.5cm]{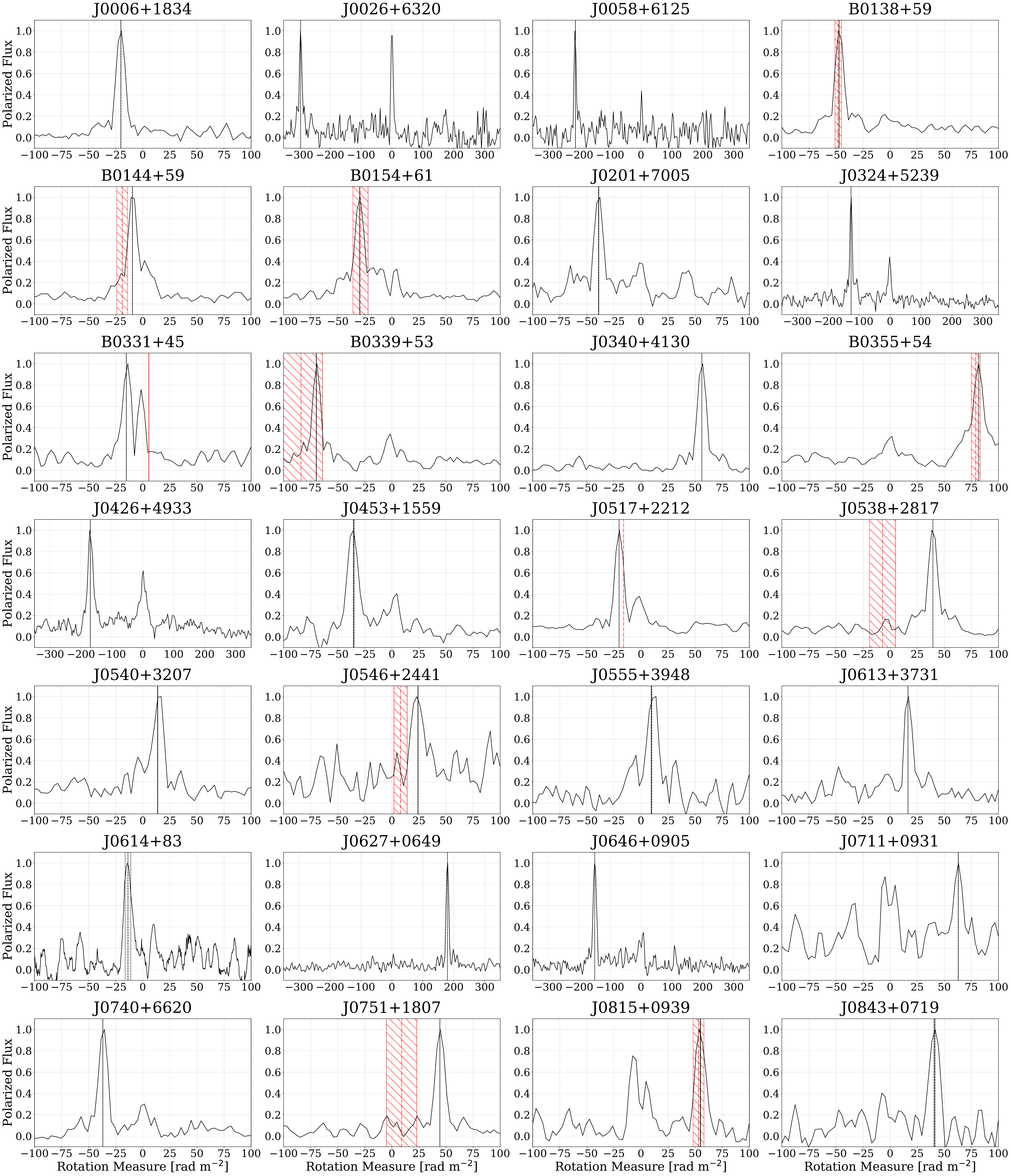}}
\caption{Faraday spectra of the 80 pulsars in this work. Refer to text for further description of the figures.}
\label{fig:plot1}
\end{figure*}

\begin{figure*}
\centering
\setlength\fboxsep{0pt}
\setlength\fboxrule{0pt}
\fbox{\includegraphics[width=17.5cm]{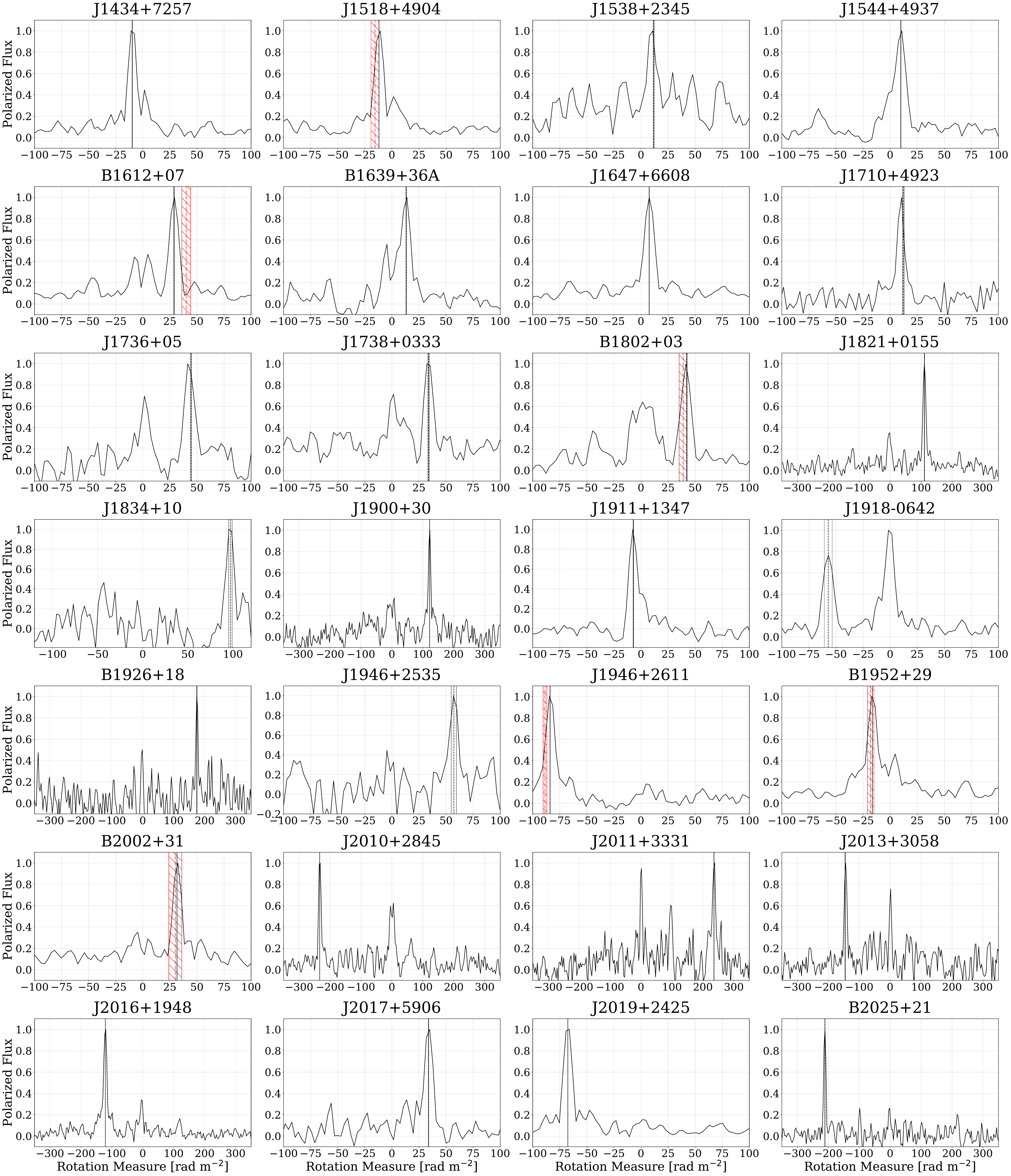}}
\caption{Continued from the previous page.} 
\label{fig:plot2}
\end{figure*}

\begin{figure*}
\centering
\setlength\fboxsep{0pt}
\setlength\fboxrule{0pt}
\fbox{\includegraphics[width=17.5cm]{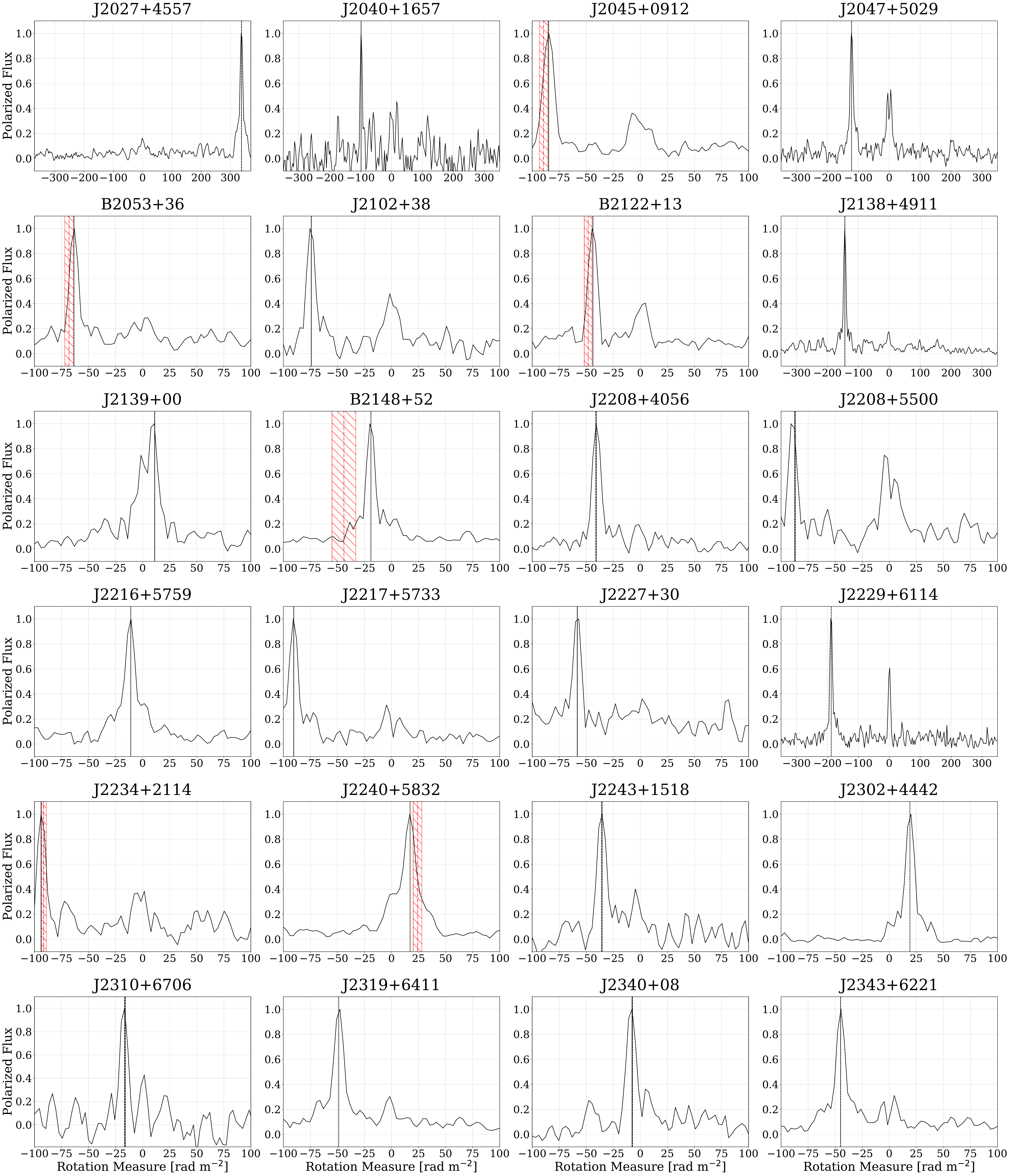}}
\caption{Continued from the previous page.} 
\label{fig:plot3}
\end{figure*}

\begin{figure*}
\centering
\setlength\fboxsep{0pt}
\setlength\fboxrule{0pt}
\fbox{\includegraphics[width=12.5cm]{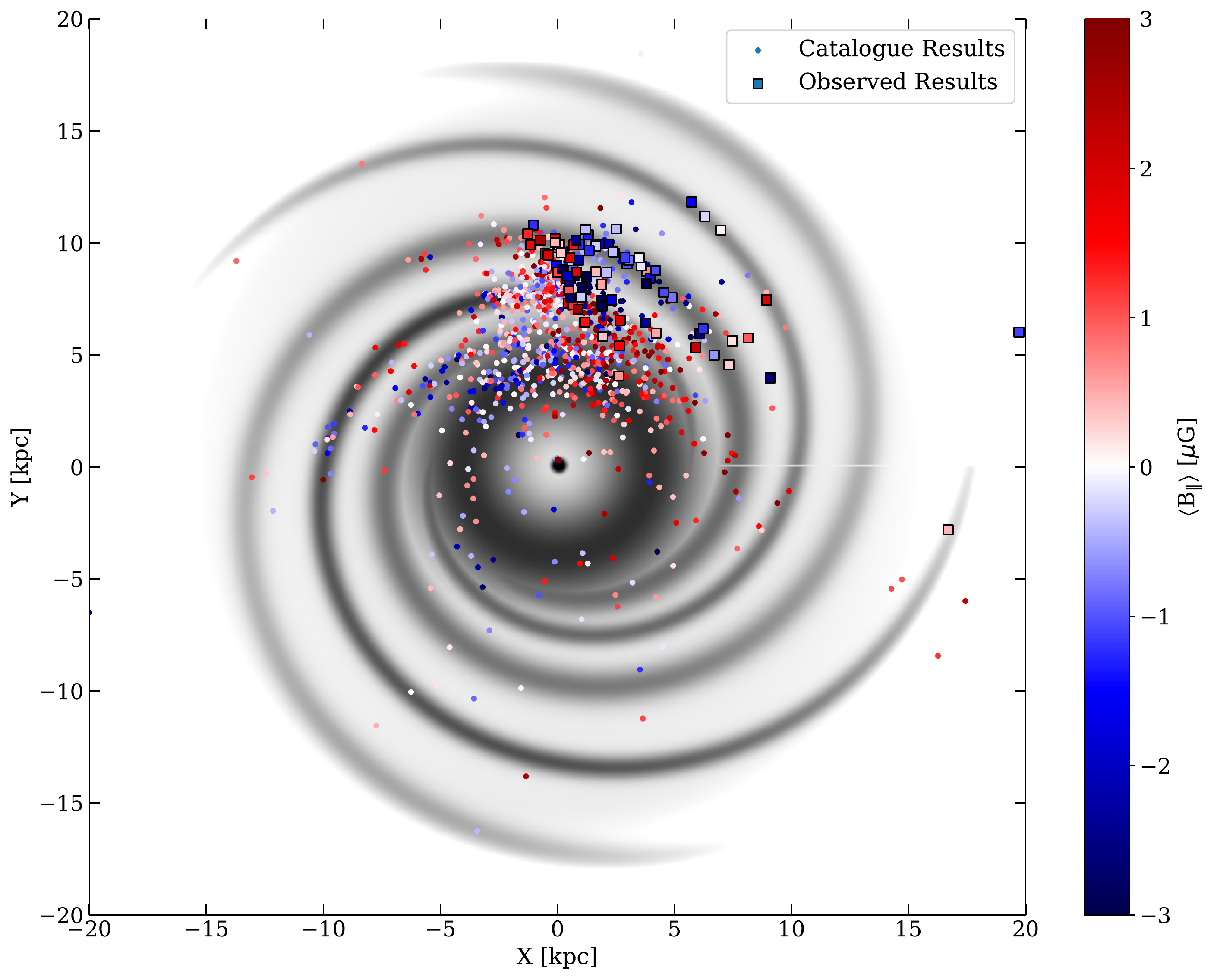}}
\caption{X-Y plane of the Milky Way, where the Sun is defined to be at (X,Y)=(0,8.3)\,kpc and the Galactic Centre at (X,Y)=(0,0)\,kpc. The pulsars with catalogue RM values are shown as circles, whereas the 80 updated RMs from this work are represented by squares. The colour scale of these symbols gives the $\left<B_{\parallel}\right>$. Pulsars with $z>|16|$\,kpc are not shown for clarity of the local distribution. The electron density in the plane of the Galaxy used in the YMW16 model is also shown in gray scale.} 
\label{fig:xyproj}
\end{figure*}

\begin{figure*}
\centering
\setlength\fboxsep{0pt}
\setlength\fboxrule{0pt}
\fbox{\includegraphics[width=12.5cm]{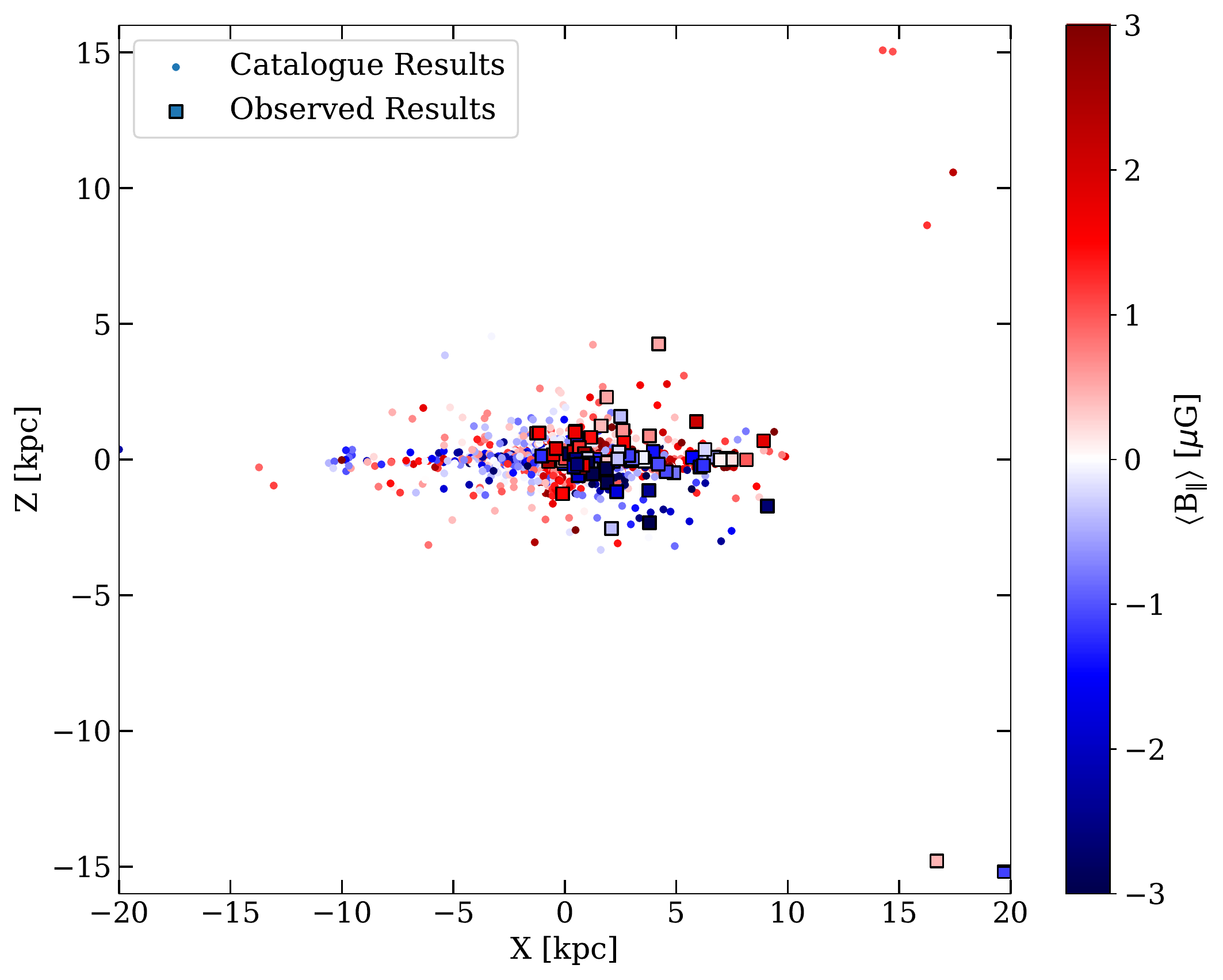}}
\caption{X-Z plane of the Milky Way, with the same plotting organization as Fig.~\ref{fig:xyproj}.} 
\label{fig:xzproj}
\end{figure*}

\begin{figure*}
\centering
\setlength\fboxsep{0pt}
\setlength\fboxrule{0pt}
\fbox{\includegraphics[width=12.5cm]{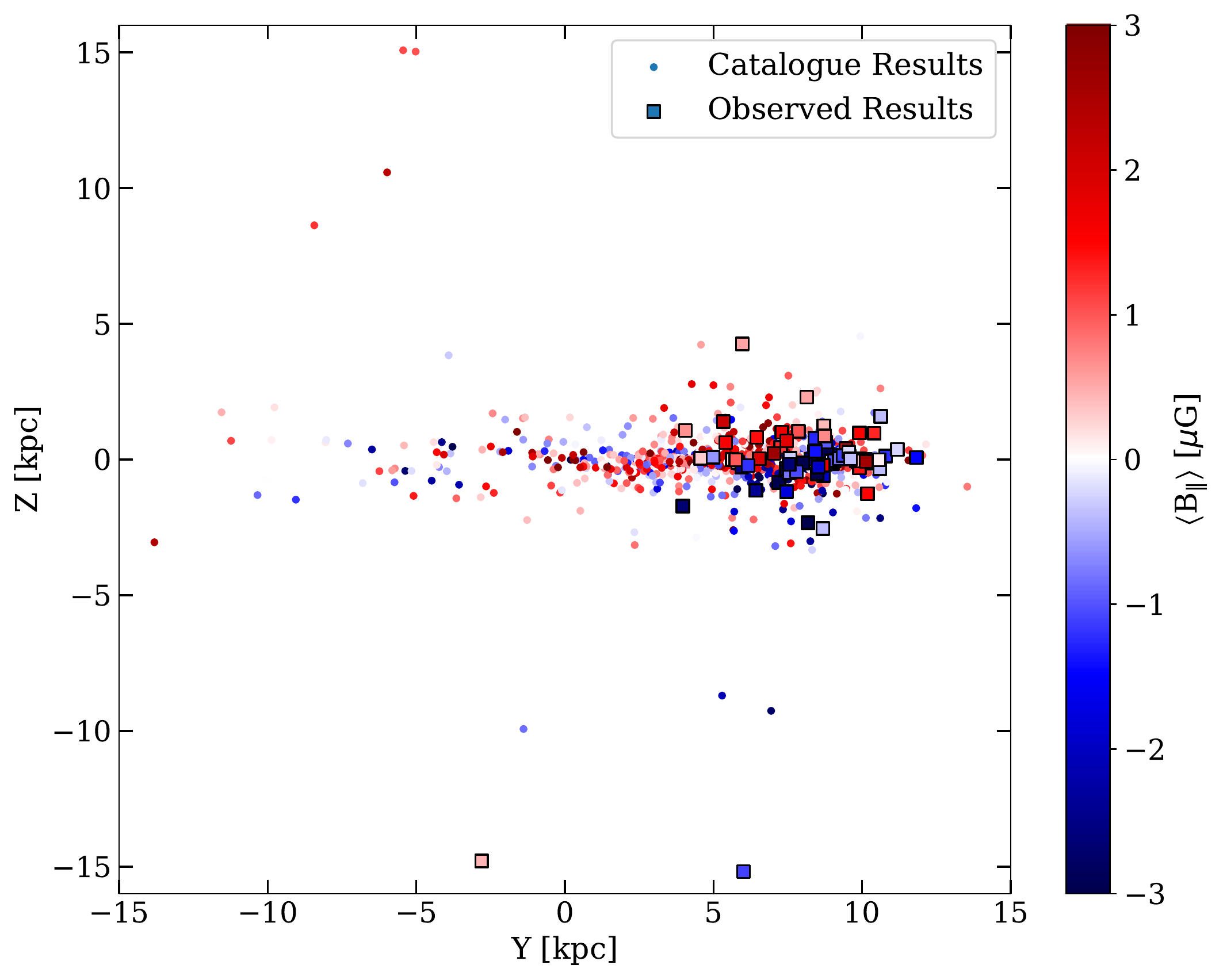}}
\caption{Y-Z plane of the Milky Way, with the same plotting organization as Fig.~\ref{fig:xyproj}. } 
\label{fig:yzproj}
\end{figure*}


\bsp	
\label{lastpage}
\end{document}